\documentclass[usenatbib]{aa}

\usepackage{graphicx}
\usepackage{amssymb}
\usepackage{epsfig}
\usepackage{natbib}

\def\chiq{$\chi^2$}

\def\B{{\em BeppoSAX }}
\def\V709{V709~Cas}
\def\igr{IGR~J00291+5934}
\def\be{\begin{equation}}
\def\ee{\end{equation}}

\begin{document}

\title{INTEGRAL Broadband X-ray spectrum of the intermediate polar V709~Cas}
\author{M. Falanga\inst{1,2}\fnmsep\thanks{\email{mfalanga@cea.fr}},
J. M. Bonnet-Bidaud\inst{1}, V. Suleimanov\inst{3,4,5} 
}

\offprints{M. Falanga}
\titlerunning{Broadband X-ray spectrum of V709~Cas}
\authorrunning{M. Falanga et al.} 

\institute{ CEA Saclay, DSM/DAPNIA/Service d'Astrophysique (CNRS FRE
  2591), F-91191, Gif sur Yvette, France
\and Unit\'e mixte de recherche Astroparticule et 
Cosmologie, 11 place Berthelot, 75005 Paris, France 
\and Kazan State University, Kremlevskaya str. 18, 420008 Kazan, Russia
\and  Astronomy Division, P.O.Box 3000, FIN-90014 University of 
  Oulu, Finland 
\and Kazan Branch of Isaac Newton Institute, Santiago, Chile
}

\abstract{We present the hard X-ray time-averaged spectrum of the
intermediate polar \V709\ observed with {\it INTEGRAL}. We performed the
observation using  data from the IBIS/ISGRI instrument in the 20--100 keV
energy band and from JEM-X at lower energy (5--20 keV). 
Using different multi-temperature and density X-ray post-shock models
we measured a improved post-shock temperature of $\sim$ 40 keV and
estimated the \V709\ mass to be 0.82$^{+0.12}_{-0.25}$ $M_{\odot}$. We
compare the resulting spectral parameters with previously reported \B\
and {\it RXTE} observations. 
}
\maketitle

\section{Introduction}
\label{sec:intro}

\V709\ (RX J0028+5917) is a member of the intermediate polars (IP)
systems, a sub-class of magnetic cataclysmic variables. This system consists of
an accreting white dwarf (WD) and a low-mass late type main sequence companion
star. The accretion onto \V709 is believed to be driven through the
Roche-lobe overflow, where the accretion flow from the secondary
proceeds towards  the IP through an accretion disk, until it reaches
the magnetospheric radius. Here the material attaches to the 
magnetic field lines and follows them almost
radially at free-fall velocity towards the magnetic poles of
the IP surface.  At some distance from this surface, the accretion flow
undergoes a strong shock, below which material settles onto the IP,
releasing X-ray as it cools by thermal bremsstrahlung processes \citep*[see
reviews][]{a73,c90,p94}. 
According to this standard model the spectra of the X-ray
post-shock emitting region has a multi-temperature and density structure
\citep[e.g.,][]{c99}.  
In order to measure the maximal shock temperature, hight energy
observatories like {\it INTEGRAL} are needed. In the most case the
temperature of the 
post-shock region of the accretion column of IPs are in the order of
$\sim$ 10--60 keV. A hard X-ray study can also be used to estimate
the IP mass by measuring the maximum temperature of the post-shock plasma
\citep{srr04}. Using the {\em INTEGRAL/RXTE} data on the
  IP V1223 Sgr \citet{r04} studied such a broadband X-ray spectrum to
  determine the shock parameters and the IP mass.
 
The X-ray V709~Cas source was discovered and identified as a IP
from the {\it ROSAT} All sky Survey \citep{hm95} and has been
extensively studied  with {\it ROSAT} \citep{hm95,m96,n99}, with a joint
{\em RXTE/BeppoSAX} X-ray observation \citep{dm01}  and using optical
spectroscopy and photometry \citep{b01,k01}, respectively.  
The pulse period was found to be 312.8 s \citep{hm95,n99} and a detailed 
spectroscopic study finally concluded to a 5.34 h orbital period 
after a previous ambiguity between 5.4 hr and 4.5 hr \citep{b01}. 

We present  here a first {\it INTEGRAL} results of a broadband spectrum
  study, from 5--100 keV, where the energies above 20 keV are based on
  hard X-ray  IBIS/ISGRI observations of \V709.

\section{Observations and Data Analysis}
\label{sec:observation}

The present dataset was obtained during the Target of Opportunity
 (ToO) A02  {\em INTEGRAL}  \citep{w03}  observation of the Cas A
 region performed  from 5--6 and 7--9 December 2004 (53345.6--53346.8
 and  53347.8--53349.8 MJD), i.e. from part of
{\em INTEGRAL} satellite revolutions  262 and 263. 
We use  data from the IBIS/ISGRI (20--100 keV) coded mask imager
 \citep{u03,lebr03} for a total exposure of 181.9 ks  and from the
 JEM-X (5--20 keV) monitor  \citep{lund03} for a total exposure time of 82 ks.
For ISGRI, the data were extracted for all pointings with a 
source position offset $\leq$ $7^{\circ}$, and for JEM-X with an 
offset $\leq$ $3.5^{\circ}$. 
The spectrometer (SPI) \citep{r03} was not used to extract the hard
 X-ray spectrum 
due to the lower sensitivity of this instrument with respect to
IBIS/ISGRI for a weak source below 100 keV. Above 90 keV, \V709
was not consistently detected in a single exposure and in the total
exposure time. 
Data reduction was performed using the standard Offline Science
Analysis (OSA) version 4.2 \citep{c03}.
The algorithms used in the spatial and spectral analysis are described
 in \citet{gold03}. 

\section{Results}
\label{sec:res}

\subsection{ISGRI Imaging}

\begin{figure}
\centerline{\epsfig{file=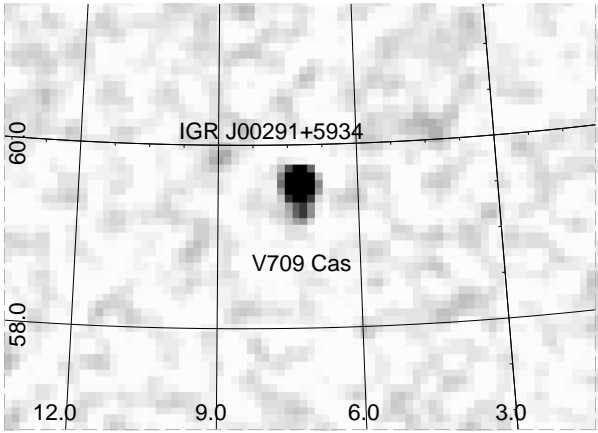,width=8cm}}
\caption
{ The 20--40 keV IBIS/ISGRI mosaicked and deconvolved sky image of the
$\sim182$ ks observation.
Image size is $\sim12^{\circ} \times 5^{\circ}$, centered at \V709
position. The pixel size is 5$'$. \V709 and IGR J00291+5934 
at a separation of 20$'$ were detected at a significance of
$\sim12.6\sigma$ and at $\sim82.8\sigma$, respectively. 
}
\label{fig:ibis_img}
\end{figure}

Fig. \ref{fig:ibis_img} shows a significance map around the source 
\V709 in the 20--40 keV energy range. Single pointings were 
deconvolved and analyzed separately, and then combined in mosaic images.
Two sources are clearly detected at a significance level of
respectively $12.6\sigma$ for \V709 and $82.8\sigma$ for IGR J00291+5934,
a newly discovered millisecond pulsar in outburst \citep{fa05}. 
In the energy band 40--80 keV, the confidence level was
$5\sigma$ for \V709 and $46\sigma$ for IGR~J00291+5934. 
At higher energies above 100 keV \V709 was not detected at a statistically
significant level either in single exposures or in the total exposure time.
To obtain precise source locations we simultaneously fitted 
the  ISGRI point spread function to the two close sources. We obtained 
a position for \V709 at $\alpha_{\rm J2000} = 00^{\rm h}28^{\rm
  m}55\fs29$ and $\delta_{\rm J2000} = 59{\degr}16\arcmin14\farcs0$. The 
position of IGR~J00291+5934 is given by $\alpha_{\rm J2000} = 00^{\rm h}29^{\rm
 m}02\fs92$ and $\delta_{\rm J2000} =
59{\degr}34\arcmin06\farcs4$. The source position offsets with respect 
to the optical catalog positions \citep{dws97,fg04} are $1\farcm4$ for
V709~Cas and $0\farcm2$ for \igr. The errors are  $1\farcm5$  and 
$0\farcm2$ for V709~Cas and \igr, respectively. These are  within 
the 90\%  confidence level assuming the source location error given by 
\citet{gros03}. The derived angular distance between the two sources 
is $\sim20'$.
Due to the fact that {\em INTEGRAL} is able to image the 
sky at high  angular resolution ($12'$ for ISGRI and $3'$ for 
JEM-X), we were able to clearly distinguish and isolate the 
high-energy fluxes from the two sources separately.
This allows us to isolate and study the X-ray emission of \V709 during the
outburst of the \igr\ pulsar. 
The lack of contamination by the pulsar outburst was verified by building 
\V709\ light curves in different energy ranges. While during the observation,
the pulsar showed a significant decay with an e-folding time of 
$\sim$ 6.6 day \citep{fa05}, the  \V709 (20-80 keV) light curve remains stable
with a constant mean counting rate of $\sim$ 0.6 $\pm$ 0.09.

\begin{figure}
\centerline{\epsfig{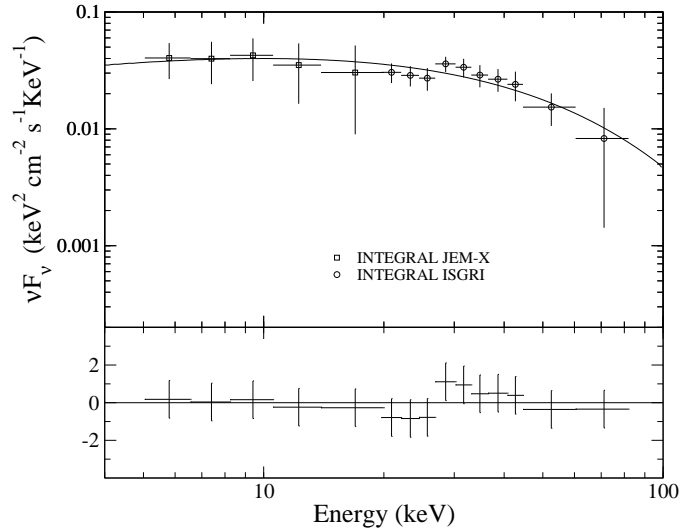}}
\caption
{The unfolded combined {\em INTEGRAL} JEM-X/ISGRI (5--100
      keV) spectrum of \V709\ with the best-fit post-shock
      model (line). Residuals between the data and model 
      are shown in the bottom panel in units of sigma.}
\label{fig:spec}
\end{figure}

\subsection{Spectral Analysis}
\label{sec:spectrum}

The spectral analysis was done using XSPEC version 11.3 \citep{a96},
combining the 20--100 keV ISGRI data with the
simultaneous 5--20 keV JEM-X data. Due to the short exposure time of
JEM-X, and therefore lower statistics, we
rebinned this data in a 5 channel energy response matrix.  
A constant factor was included in the fit to take into account the
uncertainty in the cross-calibration of the instruments.
A systematic error of 2\% was applied to the JEM-X/ISGRI spectra which
corresponds to the current uncertainty in the response matrix.
All spectral uncertainties in the results are given at a 90\%
confidence level for single parameters.

The joint JEM-X/ISGRI (5--100 keV) broadband spectrum was first
fitted with a simple optically thin thermal bremsstrahlung. 
The data fitted with {\chiq}/dof=5/12, and the plasma
temperature was found around 27 keV. In order to compare with
previously reported measurements \citep[e.g.,][]{dm01}, we also fit 
the plasma emission model {\sc Mekal}. We found a consistent plasma
temperature value with the previous thermal bremsstrahlung fit, where the
flux in the 5-100 keV energy range is 1.2$\times10^{-10}$ erg
cm$^{-2}$ s$^{-1}$ (see Table~\ref{table:fit}). During our {\it INTEGRAL}
observation the source flux in the  3--100 keV was $\sim$ 1.4 times
smaller than  that observed in July 1998 by \B and  $\sim$ 2 times
smaller by {\it RXTE} \citep{dm01,srr04}.  The obtained value of the temperature
parameter (26 keV) is significantly smaller than that obtained by
\citet{dm01} from \B\ data (42 keV) and {\it RXTE} data (36 keV) using
the same model. \citet{dm01} found  the same plasma
temperature $\sim$ 26 keV, only by including a reflection model in the fit.  We
attempted to test the hypothesis of a Compton reflection in the
spectrum including the reflection model from \citet{mz95}. 
The lack of statistics in the low energy part of the JEM-X data
prevent us to determine significant reflection parameters.    
The best fit parameters using only the  hight energy ISGRI data alone agree 
with the value from the joint JEM-X/ISGRI spectrum. This confirm the
importance of the hight energy observation to determine the post-shock
temperature above 20 keV.  

However, from the standard accretion shock model, when the matter falls
onto the WD, the X-ray post-shock emission region is expected to show a
multi-temperature and multi-density structure. To fit the
spectrum we therefore use a more physically motivated model where the X-ray
emission is determined through a density and temperature gradient along
the emission region. The broadband time-averaged
spectrum is then given by summing local bremsstrahlung spectra in the region
between the WD surface and the shock distance, $z_{0}$. We use the
same geometrical model described by
\citet{srr04}, where the observed flux is given by \citet{z90}:  

\begin{displaymath}
F_{E}  =  9.52 \times 10^{-38}   \nonumber 
\end{displaymath}
\begin{equation}
 \int^{z_0}_{R_{wd}} \biggl(\frac{\rho(z)}{\mu\, m_{\rm
    H}}\biggr)^{2}T(z)^{-1/2}
    \biggl(\frac{E}{kT(z)}\biggr)^{-0.4}\exp\biggl(-\frac{E}{kT(z)}\biggr) dz, 
\end{equation}
allowing the density profile, $\rho(z)$, and temperature profile,
$T(z)$, to vary in the post-shock region \citep*[e.g.,
  see][]{srr04,c99}. In the flux equation, $\mu=0.62$ is the mean
molecular weight of fully ionized accreting matter, $m_{\rm
    H}$ is the hydrogen mass and $k$ is the Boltzmann constant. 
The only two parameters in this model is now the WD
mass, M$_{wd}$, and the accretion rate by unit of surface $a$. 
From the best fit, using a standard 
local accretion rate of $a = 1.0$ g cm$^{-2}$ s$^{-1}$, 
we found M$_{\rm wd}$ = 0.82 M$_{\odot}$ corresponding to 
a shock temperature of 39 keV (Table~\ref{table:fit}). 
The {\chiq}\ value is comparable with one of the simple 
bremsstrahlung model, however the post-shock model represent in
more detail the physical processes in the WD emission region.
Absorption by neutral hydrogen and partial absorber have not been included
in these models since they have no significant effect
above 5 keV.  
Table \ref{table:fit} gives the best fit parameters of this column model, 
for the ISGRI and combined JEM-X/ISGRI data sets. In 
Fig. \ref{fig:spec} we present the $\nu F_{\nu}$ spectrum of the
entire observation, plotted together with the residuals in units of
$\sigma$ with  respect to the best fit post-shock model.

\subsubsection{Spectral Analysis: Models}

Different models for the structure of accretion column have 
been put forward based on assumptions of constant pressure \citep{fr02},
influence of the magnetic field \citep{wu94} or gravitational potential \citep{c99}.
In an effort to also evaluate the influence of these different assumptions, 
a phenomenological fit with a variable emission measure can be performed.
Each model is indeed characterized by a power-law type density and temperature profiles 
along the accretion column.
Assuming profiles with $(T/T_{\rm shock}) = (x/x_{\rm
  shock})^{\alpha}$  and $({\rho}/{\rho_{\rm shock}})=(x/x_{\rm
  shock})^{\beta}$, it can be shown that the emission measure 
$EM =  \int^{z_0}_{R_{wd}} \rho(z)^{2} A\, dz $ 
is defined as: 
\begin{equation}
EM=(T/T_{\rm shock})^{\Gamma}\,\; ${\rm with}$
\;\;\Gamma={\frac{(2\beta+1)}{\alpha}}.  
\end{equation}
The $EM$ therefore also follows a power-law in temperature with an
index $\Gamma$ and the column spectrum can be described with a
power-law multi-temperature plasma emission model such as the
XSPEC-{\sc cemkl}. The parameters $\alpha$ and $\beta$ have been
determined by fitting the  published profiles of the different
accretion shock structure models and are reported in table
\ref{table:fit2}.  
To determine the post-shock 
temperature, the JEM-X/ISGRI spectrum have been then fitted using {\sc cemkl} and fixing the model
dependent power-law index, $\Gamma$, and the results are reported also in
table \ref{table:fit2}.

\begin{table}[htb]
\caption{\label{table:fit} Best fit parameters of the phase-averaged
  X-ray spectra.}
\begin{flushleft}
\begin{tabular}{lll}
\hline\noalign{\smallskip}  
\noalign{\smallskip}  
Dataset & JEM-X/ISGRI & ISGRI\\
Energy range & (5--100) keV & (20--100) keV\\
\hline
Model   & \multicolumn{2}{c} {\sc Bremsstrahlung} \\
\hline\noalign{\smallskip} 
$kT_{\rm Brems}$ (keV) &  26.7$^{-6.7}_{+10.4}$ & 25.5$^{+9.3}_{-6.1}$\\ 
$\chi^{2}/{\rm dof}$   & 5.3/12 &  5.8/9 \\
$F_{\rm x}$ (erg cm$^{-2}$ s$^{-1}$)& 1.2$\times 10^{-10}$ & 0.52$\times 10^{-10}$\\ 
\hline
Model           & \multicolumn{2}{c} {\sc Mekal} \\
\hline\noalign{\smallskip} 
$kT_{\rm Max}$ (keV) &  25.7$^{+9.7}_{-6.4}$ &  24.6$^{+8.6}_{-5.8}$\\
$\chi^{2}/{\rm dof}$   & 5/12 & 5.8/9\\
$F_{\rm x}$ (erg cm$^{-2}$ s$^{-1}$) & 1.3$\times10^{-10}$ &
0.57$\times10^{-10}$\\ 
\hline
Model           & \multicolumn{2}{c} {\sc Post-Shock} \\
\hline\noalign{\smallskip} 
$M_{\rm wd}$  (M$_{\odot}$) &0.82$^{+0.12}_{-0.25}$ &0.85$^{+0.18}_{-0.15}$\\  
$kT_{\rm shock}^{\dagger}$ (keV) & 39.38 & 41.3 \\
$\chi^{2}/{\rm dof}$   & 5/12 & 6.3/9 \\
$F_{\rm x}$ (erg cm$^{-2}$ s$^{-1}$) & 1.2$\times10^{-10}$&
0.95$\times10^{-10}$\\ 
\noalign{\smallskip}  
\hline
\noalign{\smallskip}  
$^{\dagger}$ Calculated.\\
\end{tabular}
\end{flushleft}
\end{table}

\begin{table*}[htb]
\caption{\label{table:fit2} Best fit parameters from the JEM-X/ISGRI
  data in the 5--100 keV energy range.  
}
\begin{flushleft}
\begin{tabular}{llllllll}
\hline\noalign{\smallskip} 
Model & $\alpha$ & $\beta$  & $\Gamma$ & B-Field (MG) & $kT_{\rm
  shock}$ (keV) &
$\chi^{2}/{\rm dof}$ & $F_{\rm x}$ (erg cm$^{-2}$ s$^{-1}$)\\ 
\hline\noalign{\smallskip}
Frank et al. 2002 & 0.4 & -0.4 & 0.5 & -- &  39.9$^{-12.3}_{+12.1}$ &
4.91/12 &  1.36$\times10^{-10}$\\  
Wu et al. 1994  & 0.341 & -0.399 &0.592 & -- &
39.02$^{-11.8}_{+12.4}$ & 4.8/12 &  1.35$\times10^{-10}$\\ 
Wu et al. 1994 & 0.389 & -0.452 & 0.247 & 30 &
42.9$^{-13.8}_{+16.4}$ & 5.07/12 &  1.4$\times10^{-10}$\\ 
Suleimanov et al. 2005 & 0.312 & -0.433 & 0.430 & -- &
40.6$^{-12.6}_{+13.1}$ & 4.95/12 &  1.37$\times10^{-10}$\\ 
\hline\noalign{\smallskip}
\end{tabular}
\end{flushleft}
\end{table*}

\section{Conclusions} 
 \label{sec:conclusion}

\begin{figure}[t]
   \centering  
  \includegraphics[width=6.5 cm, angle=-90]{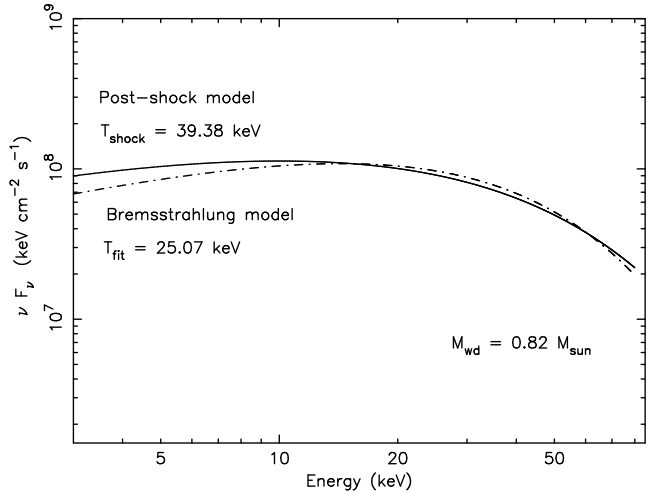}  
 \caption{The figure shows the agreement between the found value of
   the shock temperature from the post-shock model (see Sec. 3.2) with
   the best fit value using equation (1) as a Bremsstrahlung model.
}
\label{fig:model_fit}  
 \end{figure}

A first broadband spectrum result on \V709 was already studied by
\citep{dm01} using \B\ and {\em RXTE} data. We report here the first
5--100 keV high energy {\em  INTEGRAL} spectrum result of \V709.  We
found a Bremsstrahlung temperature around 26 keV which is in agreement
of a post-shock temperature around 40 keV (see Fig.
\ref{fig:model_fit}). The theoretical spectrum of the post-shock model
shown in Fig. \ref{fig:model_fit}  is calculated using equation (1), where the
bremsstrahlung spectrum is the best fit using the same equation as a
bremsstrahlung model. The Bremsstrahlung temperature is lower then
$T_{\rm shock}$ since $T_{\rm Brems}$ represent more a
weighted mean value between the temperature distribution from the IP
surface to the shock hight $z_{0}$. Using JEM-X data in the energy
band 5--20 keV and statistically significant ISGRI data in the energy
range from 20--100 keV we estimated the source mass to by
0.82$^{+0.12}_{-0.25}$ M$_{\odot}$. Using the broad band {\it RXTE}
spectrum \citet{srr04} found within the error bars the same IP mass.    
The maximum shock temperature reported in table \ref{table:fit} is
calculated from: 
\begin{displaymath}
 T_{\rm shock}= \frac{3}{8}\frac{\mu \;G \;M_{wd}\; m_{\rm
 H}}{k \;R_{wd}}. \nonumber
\end{displaymath}
Using different accretion shock structure models we found the same
post-shock temperature (see table \ref{table:fit2}).
The IP radius was calculated from the 
\citet{n72} WD mass-radius relation 
and we obtained a radius of $R_{wd} = (0.68\pm0.13) \times 10^{9}$ cm. 
The obtained estimation of the IP mass is slightly larger than that
obtained by \citet{b01}. These authors inferred the mass estimation using
optical spectroscopy, i.e. Balmer absorption lines, believed to be
from the WD photosphere. 

We estimates the accretion rate from the X-ray
luminosities of the {\it INTEGRAL} observation. For the source
distance we use the range 210--250 pc \citep{b01}.
The absorption corrected broad band X-ray flux of \V709\ in the
0.5--100 keV energy range is $F_{{\rm 0.5-100\, keV}}$ = 1.6 $\times$
10$^{-10}$ erg cm$^{-2}$ s$^{-1}$, therefore the
isotropic luminosity of one accretion colum is $L_{\rm 
  x,1}= 4\pi D^{2}$$F_{\rm 0.5-100\; keV} = 0.84-1.2 \times 10^{33}$
erg s$^{-1}$.
From the Bremsstrahlung model the emission
measure of one emitting accretion colum is $E_{m,1}$ = $\int N_{e}^{2}$
dV = 4.72 $\times$ 10$^{12}$ 4$\pi D^{2}$ cm$^{-3}$, where $N_{e}$ is
the number density of the electron density in a hot emitting
plasma.  The volume emission measure therefore is $E_{m,1} \approx$ 2.5--3.5
$\times 10^{55}$ cm$^{-3}$. Consistent with the value found by \citet{dm01}. 
As the X-ray emission is the 
major channel of energy losses in the post-shock regio we derived the
mass accretion rate $\dot M$
\begin{displaymath}
\dot M \approx \frac{3}{8}\frac{\mu\; m_{\rm H}\,G\,2\,L_{\rm
    x,1}}{kT_{\rm shock}}. \nonumber 
\end{displaymath}
The factor 2 stands in order to account for two emitting magnetic
poles of the IP. Substituting measured values we obtained $\dot M
= 0.82-1.2 \times 10^{16}$ g s$^{-1}$. 
Using $m_{wd}$ = 0.82 in units of the solar mass, $R_{wd}= 0.68 \times 10^{9}$ cm and
the $\dot M$ we evaluated the height of the standing shock $h_{s}$ and
the number density of electrons in the post-shock region $N_{e}$ following
\citet{warner95}.    
\begin{displaymath}
N_{e} = 3.1 \times 10^{15} \dot M_{16}
\biggr(\frac{m_{wd}}{M_{\odot}}\;\biggl)^{1/2} R_{9}^{-3/2} f_{-3}^{-1} \;\;
\mbox{cm$^{-3}$} \nonumber 
\end{displaymath}
\begin{displaymath}
h_{s} = 9.6 \times 10^{7}\; m_{wd} \biggr(\frac{N_{e}}{10^{16} {\rm cm^{-3}}}\biggl)^{-1} R_{9}^{-1}\; \mbox{cm}, \nonumber 
\end{displaymath}
where $\dot M_{16}$ is the mass accretion rate of 10$^{16}$ g
s$^{-1}$, $R_{9}$ is the IP radius in the units of $10^{9}$ cm,  and
$f$ is the fraction of the IP surface, occupied by two 
accretion columns. We derived $N_{e}= 4.1-5.8\times
10^{15}f_{-3}^{-1}$ cm$^{-3}$ and $h_{s} =  0.21-0.28 \;
f_{-3}$ $R_{wd}$, where $f_{-3}=10^{-3}f$ is the fraction of the surface of
the IP, occupied by two accretion columns in units of $10^{-3}$. 
The derived parameters are similar to those obtained by \citet[][]{dm01,b01}.

\acknowledgements
MF acknowledges the French Space Agency and CNRS  for financial
support. VS was partially supported by the Academy of Finland exchange grant. 
MF and VS are grateful to the NORDITA Nordic project on
High Energy Astrophysics and the University of Oulu where part of this
work was done.

\end{document}